\documentclass[twocolumn,aps,pra,superscriptaddress,footinbib,amsmath,amssymb,showpacs]{revtex4-1} %
\usepackage{graphicx}
\usepackage{bm}
\usepackage{mathrsfs}
\usepackage{mathtools}
\usepackage[colorlinks,	linkcolor=blue,anchorcolor=blue,citecolor=blue,bookmarksnumbered]{hyperref}

\makeatother

\begin{document}
\title{Sensing weak anharmonicities with a passive-active anti-PT symmetric
system}
\author{Ya-Wei Zeng}
\affiliation{Fujian Key Laboratory of Quantum Information and Quantum Optics and
Department of Physics, Fuzhou University, Fuzhou 350116, China}
\author{Wei-Xin Chen}
\affiliation{Fujian Key Laboratory of Quantum Information and Quantum Optics and
Department of Physics, Fuzhou University, Fuzhou 350116, China}
\author{Tian-Le Yang}
\affiliation{Fujian Key Laboratory of Quantum Information and Quantum Optics and
Department of Physics, Fuzhou University, Fuzhou 350116, China}
\author{Wan-Jun Su}
\affiliation{Fujian Key Laboratory of Quantum Information and Quantum Optics and
Department of Physics, Fuzhou University, Fuzhou 350116, China}
\author{Huaizhi Wu} 
\affiliation{Fujian Key Laboratory of Quantum Information and Quantum Optics and
Department of Physics, Fuzhou University, Fuzhou 350116, China}
\begin{abstract}
We propose a scheme for enhanced sensing of weak anharmonicities based
on a three-mode anti-parity-time (anti-PT) symmetric cavity-magnon-waveguide
system. By tuning the optical gain to the active cavity mode, the
linewidth suppression point for the anti-PT symmetric Hamiltonian
can be flexibly controlled even when the two dissipative magnonic
modes experience strong intrinsic decay. This essential characteristic
is utilized for detecting weak nonlinearities in both the cavity and
magnonic modes, with both demonstrating similar high levels of sensitivity.
Moreover, the sensitivity can be greatly improved with a detuned laser
drive. Based on the integrated passive-active three-mode anti-PT symmetric
system, the sensing scheme can be generalized to various physical
systems with anharmonicities.
\end{abstract}
\maketitle

\section{INTRODUCTION}

Sensing has attracted widespread interest across a broad range of
fields, including optics \citep{chen2017exceptional,hodaei2017enhanced,RevModPhys.86.1391},
magnon-cavity hybrid systems \citep{PhysRevLett.125.147202}, and
electrical circuit resonators \citep{PhysRevApplied.13.064022,2018Observation}.
In quantum mechanics, the Hermicity of the Hamiltonian and its real
eigenvalues in closed systems stand as fundamental postulates. Regarding
sensing within conventional Hermitian systems, the eigenvalue splitting,
which refers to the energy splitting between two levels, induced by
a linear perturbation $\epsilon$ is directly proportional to the
magnitude of $\epsilon$. Mathematically, the sensitivity of a Hermitian
system to a linear perturbation can be expressed as a functional dependence
$\Delta\omega\propto\epsilon$. In recent years, the enhancement of
sensitivity by harnessing non-Hermitian degeneracies and exceptional
points (EPs) has been the subject of intensive theoretical and experimental
investigations \citep{PhysRevLett.123.213901,PhysRevLett.117.110802,PhysRevApplied.5.064018,PhysRevResearch.4.013131,PhysRevA.93.033809,PhysRevApplied.16.044016,Li:23,PhysRevLett.125.240506,Felski2025}.
Parity-time (PT) symmetric systems, a unique subclass of non-Hermitian
systems, are defined by the invariance of their Hamiltonian under
the combined transformation of the PT operator \citep{PhysRevLett.80.5243,PhysRevLett.89.270401,bender2002generalized,ozdemir2019parity,Roy2025,Wang2025,Wang2025a,Wang2025b},
satisfy a specific commutation relation formulated as $[H,PT]=0$.
When subjected to a linear perturbation, the resulting eigenvalue
splitting follows a scaling law of $\epsilon^{1/N}$, where the PT
symmetric system's $N$ eigenvalues and their corresponding eigenvectors
converge. Consequently, for sufficiently small perturbations, the
eigenvalue splitting behavior at EPs is significantly more sensitive
than that in Hermitian systems. This characteristic has already found
applications in many practical sensing scenarios \citep{https://doi.org/10.1002/qute.202300350,PhysRevA.105.043505,zeng2019enhanced}.

Sensing schemes based on anti-parity-time (anti-PT) symmetric systems
have also garnered considerable attention \citep{PhysRevLett.129.273601,li2019anti,zhang2020breaking,PhysRevB.105.064405,PhysRevA.104.012218,PhysRevA.104.L031503,PhysRevA.107.033507,PhysRevA.96.053845,Qin:21,PhysRevLett.128.173602,PhysRevLett.124.053901,Li2025,Jahangiri2025,Ren2025,Yang2024,Li2024a}.
For anti-PT symmetric systems, the Hamiltonian exhibits an anti-commutation
relationship with the PT operator, which can be mathematically expressed
as $\{PT,H\}=0$. The anti-PT symmetric systems can be applied not
only for the sensing of linear perturbations \citep{zhang2020breaking},
but also, remarkably, for efficient sensing of nonlinear perturbations
\citep{PhysRevLett.126.180401}. The latter is built upon a dissipatively
coupled two-mode system \citep{PhysRevLett.126.180401} with vacuum-induced
coherence (VIC) between the two modes \citep{PhysRevA.85.011401,PhysRevB.86.205315,PhysRevLett.84.5500}.
The VIC can induce linewidth suppression, a singularity point which
can be used for enhanced sensing of weak magnetic nonlinearities.
This paradigm has also been generalized to integrated cavity-waveguide
optomechanical systems, where the joint effect of optical bistability
and linewidth suppression can greatly enhance the sensitivity of detecting
optomechanically induced nonlinearities \citep{Liu:23}. Nevertheless,
the practical implementation of such two-mode anti-PT symmetric systems
is hindered by inherent drawbacks. Specifically, there exist unavoidable
intrinsic losses for the coupled subsystems \citep{PhysRevB.99.134426,PhysRevLett.113.083603,PhysRevLett.113.156401},
which would lead to a strong quenching in the magnon response to the
nonlinearities and thus its sensitivity to anharmonicities \citep{PhysRevLett.126.180401}.

In order to effectively circumvent the drawbacks inherent in two-mode
anti-PT symmetric systems, in this paper, we explore sensing of Kerr
nonlinearities by using a passive-active anti-PT symmetric system,
comprising one active cavity mode and two dissipative magnonic modes.
Intriguingly, the VIC-induced linewidth suppression, typically associated
with intrinsically lossless systems, can still be achieved and actively
controlled through optical gain---even when the two magnon modes
experience significant individual losses, with decay rates comparable
to half the dissipative coupling strength. Note that the linewidth
suppression point functions as the EP in PT-symmetric systems, and
can be used for enhanced sensitivity to weak Kerr nonlinearity in
both the cavity mode and the magnonic modes. The anti-PT symmetric
system keeps away from optical bistability with resonant driving,
corresponding to the linewidth suppression point. However, by introducing
an appropriate laser detuning to the cavity mode, the sensitivity
to the weak nonlinearities can be increased by about fifty percent.
Thus, the three-mode passive-active anti-PT symmetric configuration
exhibits notable advantages compared to conventional two-mode systems,
and may find potential applications in a wide class of systems, including
waveguide cavity optomechanical systems and hybrid cavity magnonic
systems.

The paper is structured as follows. Section \ref{sectionII} is dedicated
to the introduction of our physical model and the detailed elaboration
of the eigenvalues of the anti-PT symmetric Hamiltonian. Section \ref{sectionIII}
showcases the enhanced sensing capabilities for weak nonlinearities
in both the cavity mode (Sec. \ref{sectionIII A}) and the magnonic
modes (Sec. \ref{sectionIII B}). Section \ref{section IV} delves
into the practical implementation aspects of the physical model, and
then gives the conclusion of this work. In Appendix \ref{Appendix A},
we present the stability criterion for our three-mode anti-PT symmetric
system. Appendix \ref{Appendix B} elaborates on the system dynamics
when a non-zero laser detuning to the cavity mode is taken into account.

\section{\label{sectionII}The anti-PT symmetric system with an active mode}
\begin{figure}
\includegraphics[width=0.9\columnwidth]{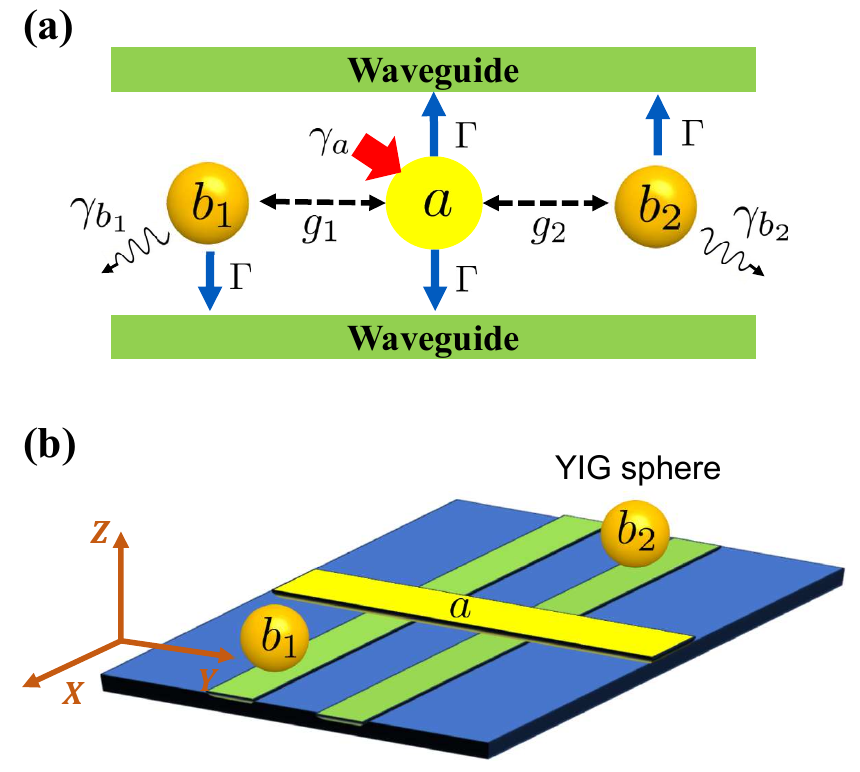}
\caption{\label{Model} (a) Theoretical model and (b) Schematic of the cavity-magnon-waveguide
setup. The microwave cavity running transverse to the waveguide interacts
with two YIG spheres via the transmission line. $g_{1}$ and $g_{2}$
are the coherent coupling strengths between the active cavity mode
$a$ and the two magnon modes $b_{1}$ and $b_{2}$, respectively.
The magnon modes $b_{1}$ and $b_{2}$ also dissipatively couple to
the cavity via the waveguides, with the rates $\Gamma$ being identical
to the cavity-waveguide coupling. $\gamma_{b_{1}}$ and $\gamma_{b_{2}}$
are the damping rates of $b_{1}$ and $b_{2}$, and $\gamma_{a}=\kappa_{-}-\kappa_{+}$
denotes the effective gain of the cavity mode $a$ (i.e., the net
rate defined by the difference between cavity loss and incoherent
gain).}
\end{figure}

We consider a three-mode cavity-magnon-waveguide system as schematically
illustrated in Figs. \ref{Model} (a) and \ref{Model}(b). A microwave
cavity (mode $a$) is dissipatively coupled to two yttrium iron garnet
(YIG) spheres --- each hosting a magnon mode $b_{j}$ $(j=1,2$)
--- via two separate waveguides, with coupling strengths $\Gamma$
between mode $a$ and each $b_{j}$ $(j=1,2$). Due to spatial overlap,
the cavity mode $a$ may also coherently couple to magnon mode $b_{j}$
with weak strength $g_{j}$. The cavity mode $a$ is externally driven
by a laser operating at a frequency of $\omega_{d}$ with the strength
$\Omega$. In addition, the cavity mode $a$ and the magnon modes
$b_{j}$ may exhibit Kerr-type anharmonicity, arising either from
the nonlinear response of the electric polarization in optical resonators
\citep{Haroche2006,PhysRevA.76.042319,ZARERAMESHTI20221,RevModPhys.93.025005}
or from the intrinsic nonlinear magnetization dynamics in magnetic
systems \citep{PhysRevLett.130.033601,PhysRevB.94.224410,PhysRevLett.120.057202}.
The Hamiltonian of the system can be effectively described by 
\begin{eqnarray}
H & = & \sum_{j=1,2}[\omega_{j}b_{j}^{\dagger}b_{j}+g_{j}(b_{j}^{\dagger}a+b_{j}a^{\dagger})]+\omega_{a}a^{\dagger}a\nonumber \\
 &  & +\sum_{j=1,2}U_{b_{j}}b_{j}^{\dagger2}b_{j}^{2}+U_{a}a^{\dagger2}a^{2}\nonumber \\
 &  & +i\Omega(a^{\dagger}e^{-i\omega_{d}t}-ae^{i\omega_{d}t}),\label{eq:Hamiltonian}
\end{eqnarray}
where $\omega_{j}$ ($\omega_{a}$) are the resonance frequencies
of mode $b_{j}$ ($a$). The parameters $U_{b_{j}}$ characterize
the intrinsic magnon anharmonicity of the YIG modes\textbf{ $b_{j}$}
\citep{PhysRevB.94.224410}; while $U_{a}$ denotes the cavity\textquoteright s
self-Kerr nonlinearity, which can arise naturally, for example, through
coupling to a nonlinear qubit \citep{RevModPhys.93.025005}.

\begin{figure*}
\includegraphics[width=2\columnwidth]{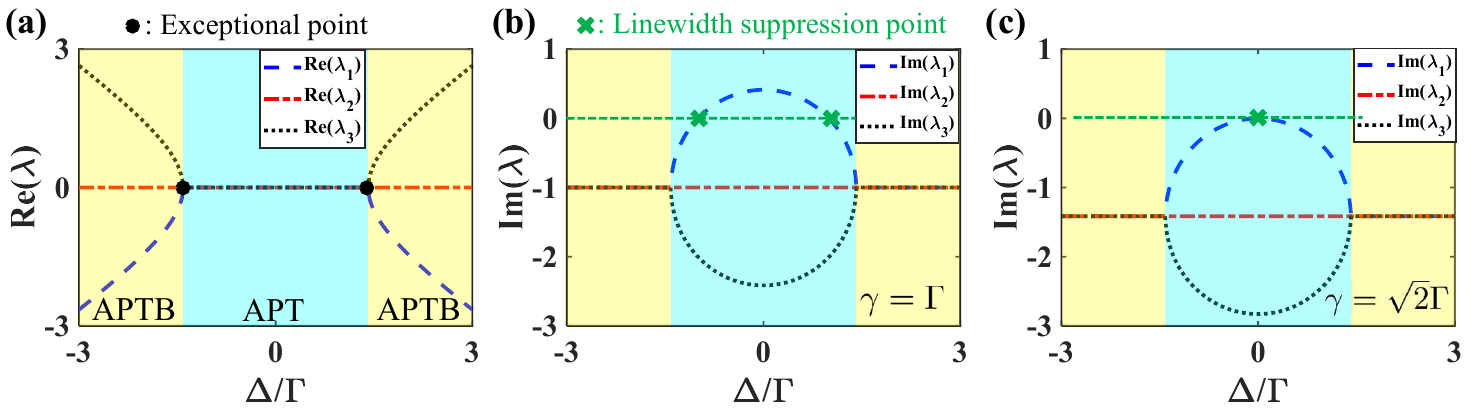}\caption{\label{Eigenvalues} (a) Real parts (eigenfrequencies) and (b), (c)
imaginary parts (linewidths) of the eigenvalues for the anti-PT symmetric
system (with $g=0$) plotted against $\Delta$. The EPs appear at
$\Delta=\pm\sqrt{2}\Gamma$. APT denotes the anti-PT symmetric phase,
and APTB represents the anti-PT symmetry broken phase. Panels (b)
and (c), show the imaginary parts of the eigenvalues under the conditions
of $\gamma=\Gamma$ and $\gamma=\sqrt{2}\Gamma$, respectively. The
green dashed line is used to screen out the parameter conditions when
the imaginary part of the eigenvalue is zero, which are marked with
green crosses.}
\end{figure*}

We introduce the Lindblad master equation to account for dissipative
processes mediated by the environment, including both reservoir-induced
coupling through the common waveguide and spontaneous decay channels
arising from thermal baths \citep{Galindo2004QuantumNA}. When the
localized modes $b_{1(2)}$ and $a$ are coupled to a one-dimensional
waveguide with a finite separation, adiabatic elimination of the waveguide
degrees of freedom under the Markov approximation yields a master
equation describing the resulting coherent and dissipative interactions.
In addition, incoherent pumping may be applied to the cavity mode
$a$ to generate optical gain, a mechanism well established in microwave
QED systems, including inverted qubits \citep{Quijandria2018} and
engineered reservoirs \citep{Metelmann2015}. Incorporating these
decay and gain channels leads to the driven-dissipative dynamics of
the full three-mode system as described by the master equation for
the density matrix $\rho$:
\begin{eqnarray}
\frac{d\rho}{dt} & = & -\frac{i}{\hbar}[H,\rho]+\gamma_{b_{1}}\mathcal{L}[b_{1}]\rho+\gamma_{b_{2}}\mathcal{L}[b_{2}]\rho\nonumber \\
 &  & +\text{\ensuremath{\kappa_{-}}}\mathcal{L}[a]\rho+\text{\ensuremath{\kappa_{+}}}\mathcal{L}[a^{\dagger}]\rho+2\Gamma\mathcal{L}[m_{1}]\rho+2\Gamma\mathcal{L}[m_{2}]\rho,\nonumber \\
\label{eq:Master equation}
\end{eqnarray}
where $\mathcal{L}$ is the Liouvillian operator given by $\mathcal{L}[\sigma]\rho=2\sigma\rho\sigma^{\dagger}-\sigma^{\dagger}\sigma\rho-\rho\sigma^{\dagger}\sigma$
with $\sigma=a,a^{\dagger},b_{j}$. The term $\mathcal{L}[b_{1(2)}]\rho$
describes spontaneous decay of the magnonic modes with the rate $\gamma_{b_{1(2)}}$.
The terms $\kappa_{-}\mathcal{L}[a]\rho+\kappa_{+}\mathcal{L}[a^{\dagger}]\rho$
represent an open cavity with both loss (of the rate $\kappa_{-}$)
and incoherent gain, where photons are injected at rate $\kappa_{+}$
(e.g., through optical pumping or coupling to an inverted atomic ensemble)
\citep{Metelmann2015,PhysRevB.97.241301,PhysRevLett.132.183803}.
The waveguide-mediated dissipator takes the form $\mathcal{L}[m_{j}]\rho$,
where the jump operators $m_{j}$ ($j=1,2$) is a linear superposition
of the annihilation operators $b_{j}$ and $a$, i.e. $m_{j}=v_{j}b_{j}+u_{j}e^{i\Phi_{j}}a$.
Here $\Phi_{j}$ denote the propagation phase acquired between the
mode positions, and the coefficients $v_{j}$ and $u_{j}$ characterize
their respective coupling strengths to the travelling-wave reservoir
\citep{Metelmann2015}. The associated decay rate into the waveguide
is the standard Fermi--Golden-Rule rate $\Gamma$, which represents
the external (radiative) damping depending on the vacuum coupling
strength, the local photonic density of states (or the group velocity
of the waveguide mode). When the cavity and magnon modes couple identically
to the common reservoir and the propagation phases are tuned to $\Phi=\Phi_{1}=\Phi_{2}=2k\pi$
(with integer $k$), the dissipative coupling leads to symmetric optical--magnonic
supermodes of the form $m_{j}=(b_{j}+a)/\sqrt{2}$. 

To treat the nonlinear term analytically, we adopt the mean-field
approximation, i.e. $\langle o_{1}o_{2}\rangle=\langle o_{1}\rangle\langle o_{2}\rangle$
for any two operators. We further assume that the system operates
in a zero-temperature environment with white Gaussian noise. By choosing
a proper rotating frame with respect to the driving frequency $\omega_{d}$,
and denoting $\alpha(t)=\langle a(t)\rangle$ ($\beta_{j}=\langle b_{j}(t)\rangle$),
the simplified form of the evolution equation is expressed as 
\begin{eqnarray}
\left(\begin{array}{c}
\dot{\beta_{1}}\\
\dot{\alpha}\\
\dot{\beta_{2}}
\end{array}\right) & = & -i\mathscr{\mathcal{H}}\left(\begin{array}{c}
\beta_{1}\\
\alpha\\
\beta_{2}
\end{array}\right)-2i\mathscr{\mathcal{R}}\left(\begin{array}{c}
\beta_{1}\\
\alpha\\
\beta_{2}
\end{array}\right)+\Omega\left(\begin{array}{c}
0\\
1\\
0
\end{array}\right),\label{eq:HeisenbergOM}
\end{eqnarray}
with 
\begin{equation}
\mathscr{\mathcal{R}}=\left(\begin{array}{ccc}
U_{b_{1}}\beta_{1}^{*}\beta_{1} & 0 & 0\\
0 & U_{a}\alpha^{*}\alpha & 0\\
0 & 0 & U_{b_{2}}\beta_{2}^{*}\beta_{2}
\end{array}\right)
\end{equation}
and
\begin{eqnarray}
\mathscr{\mathcal{H}} & = & \left(\begin{array}{ccc}
\tilde{\Delta}_{1}-i\Gamma & g_{1}-i\Gamma & 0\\
g_{1}-i\Gamma & \Delta_{a}-i(\gamma_{a}+2\Gamma) & g_{2}-i\Gamma\\
0 & g_{2}-i\Gamma & \tilde{\Delta}_{2}-i\Gamma
\end{array}\right),\label{eq:Hamiltonian matrix}
\end{eqnarray}
where $\tilde{\Delta}_{j}=\Delta_{j}-i\gamma_{b_{j}}$ with $\Delta_{j}=\omega_{j}-\omega_{d}$,
$\Delta_{a}=\omega_{a}-\omega_{d}$, and $\gamma_{a}\eqqcolon\kappa_{-}-\kappa_{+}$
denotes the effective damping rate of the cavity mode, with positive
(negative) values corresponding to net loss (gain).

We begin with an analysis of the linear dynamics by setting $U_{a}=U_{b_{j}}=0$.
Moreover, we assume that $\Delta_{a}=0$, $\Delta=\Delta_{1}=-\Delta_{2}$,
$\gamma=\gamma_{b_{j}}+\Gamma=\gamma_{a}+2\Gamma$, then the eigenvalues
of $\mathcal{H}$ read 
\begin{eqnarray}
\lambda_{1} & = & -i\left(\gamma-\sqrt{2\Gamma^{2}+4ig\Gamma-2g^{2}-\Delta^{2}}\right),\nonumber \\
\lambda_{2} & = & -i\gamma,\label{eq:Eigenvalues(TravelingPhase)}\\
\lambda_{3} & = & -i\left(\gamma+\sqrt{2\Gamma^{2}+4ig\Gamma-2g^{2}-\Delta^{2}}\right),\nonumber 
\end{eqnarray}
where we have kept the coherent coupling strength $g_{1}=g_{2}=g$
for later discussions. In the absence of spatial overlap between the
cavity and magnon modes, the direct couplings between them are negligible,
i.e. $g_{1}=g_{2}=0$. Then, $\mathcal{H}$ reduces to a three-mode
anti-PT symmetric Hamiltonian given by 
\begin{eqnarray}
\mathscr{\mathcal{H}} & = & \left(\begin{array}{ccc}
\Delta-i\gamma & -i\Gamma & 0\\
-i\Gamma & -i\gamma & -i\Gamma\\
0 & -i\Gamma & -\Delta-i\gamma
\end{array}\right),\label{eq:Anti-PT symmetric matrix}
\end{eqnarray}
which obeys $(PT)\mathcal{H}(PT)^{-1}=-\mathcal{H}$, with the parity
operator ($P$) exchanging the first and third modes, and keeping
the second mode fixed. We show the eigenvalues with $g=0$ in Fig.
\ref{Eigenvalues}. The blue area in the figure represents the anti-PT
symmetric region (APT), while the yellow areas represent the anti-PT
symmetry broken regions (APTB). The EPs occur at $\Delta=\pm\sqrt{2}\Gamma$
where the three eigenvalues coalesce and become degenerate. Moreover,
linewidth suppression arises when both the real and imaginary parts
of an eigenvalue simultaneously approach zero, and appears in the
anti-PT symmetric phase ($|\Delta|<\sqrt{2}\Gamma$) where $E_{p}\equiv\Delta^{2}+\gamma^{2}-2\Gamma^{2}\rightarrow0$.

\begin{figure*}
\includegraphics[width=1\textwidth]{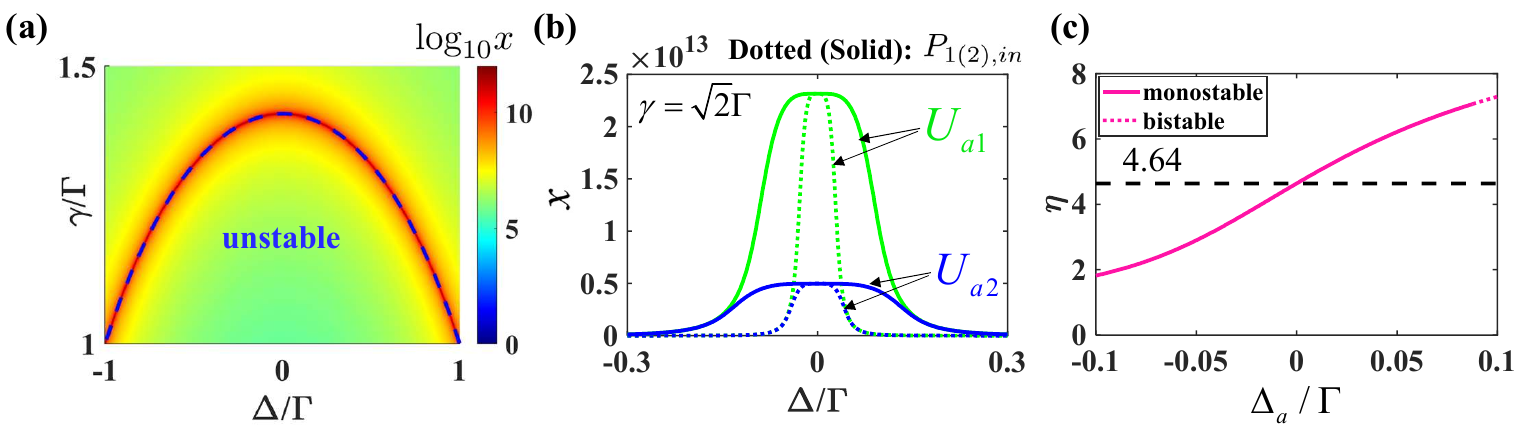}\caption{\label{Response1} (a) The rescaled cavity response $\textrm{log}_{10}x$
plotted against $\Delta$ and $\gamma$ for $U_{a}/2\pi=1\textrm{ nHz}$
and $\Delta_{a}=0$. The blue dashed line delineates the boundary
of the unstable region and corresponds to the set of linewidth suppression
points. (b) The cavity responses plotted against $\Delta$ with different
nonlinearity strengths $U_{a1}/2\pi=0.1\textrm{ nHz}$ and $U_{a2}/2\pi=1\textrm{ nHz}$
and two different drive powers. Here we have set $\gamma=\sqrt{2}\Gamma$,
$\Delta_{a}=0$, $P_{1,in}=8$ $\mu$W, and $P_{2,in}=8$ mW. At $\Delta=0$,
the peak in the cavity response arises from linewidth suppression.
For comparison, the green- and blue-dotted curves have been scaled
up by 10. (c) The ratio $\eta$ as a function of $\Delta_{a}$, at
the drive power of $8\textrm{ mW}$ and $\Delta=0$ with $U_{a1}=0.1\textrm{ \ensuremath{\mu}Hz}$
and $U_{a2}=1\textrm{ \ensuremath{\mu}Hz}$. The monostable regime
and bistable regime are denoted by the pink solid line and dotted
line, respectively. The horizontal dashed line represents the response
enhancement factor of 4.64, corresponding to the sensitivity at the
linewidth suppression condition. Other parameters are: $\Gamma/2\pi=1\textrm{ MHz}$,
$\kappa_{-}=0.05\Gamma$, $\lambda_{d}=1550\textrm{ nm}$, $\Omega=\sqrt{P_{j,in}\kappa_{-}/\hbar\omega_{d}}$.}
\end{figure*}

For the three-mode system discussed here, the condition $E_{p}\approx0$
implies that the linewidth suppression signature is geometrically
governed by an elliptic curve in the ($\Delta$, $\gamma$) parameter
space (normalized by $\Gamma$), which also serves as the stability
boundary (see Appendix A). Remarkably, the linewidth suppression effect
persists even when both modes $b_{1}$ and $b_{2}$ remain lossy,
due to the presence of gain in the cavity mode. As can be seen in
Fig. \ref{Eigenvalues}(b), when the decay rates of the magnon modes
are vanishing, i.e. $\gamma_{b_{j}}=0$ (or $\gamma=\Gamma$), the
singularities arise at $\Delta=\pm\Gamma$ for the optical gain $-\gamma_{a}=\Gamma$.
For the special case of $\gamma=\sqrt{2}\Gamma$, there exists only
a unique singularity point with zero linewidth, as shown in Fig. \ref{Eigenvalues}(c).
In this case, the decay rate of the magnon modes is explicitly given
by $\gamma_{b_{j}}=(\sqrt{2}-1)\Gamma$, and the effective gain of
the cavity mode is $-\gamma_{a}=(2-\sqrt{2})\Gamma$. In general,
two singularities associated with linewidth suppression can be observed
at the detuning $\Delta=\pm\sqrt{\Gamma^{2}-\gamma_{b_{j}}^{2}-2\Gamma\gamma_{b_{j}}}$
, provided that the magnon decay rate lies within the range $0<\gamma_{b_{j}}<(\sqrt{2}-1)\Gamma$.
This highlights a clear advantage compared to two-mode systems, in
which linewidth suppression is limited to cases where none of the
modes suffers spontaneous loss to its independent surrounding \citep{PhysRevLett.126.180401}.

\section{\label{sectionIII}Enhanced sensing to anharmonicities due to linewidth
suppression}

For sensing of weak anharmonicities, one can simply consider the resonant
condition $\Delta=0$ (where the cavity and magnon modes are resonant),
and tailor the magnon--waveguide coupling to meet $\gamma_{b_{j}}=(\sqrt{2}-1)\Gamma$,
which allows the resonant cavity drive to produce a large and highly
nonlinear cavity (or magnon) response, see the details later. In practice,
however, when the two magnon modes are non-degenerate $\Delta\neq0$,
one can tune the optical gain so that the operating point falls inside
the stable region, near linewidth-suppression point.

With the appropriate choice of the driving and pumping strengths,
the system is in the stable regime. The steady-state solutions for
the mean values $\beta_{1}$, $\alpha$ and $\beta_{2}$ of the corresponding
modes $b_{1}$, $a$ and $b_{2}$ are
\begin{eqnarray}
\beta_{1} & = & \frac{-\Gamma(-i\Delta+\gamma)}{(\Delta^{2}+\gamma^{2}-2\Gamma^{2})\gamma}\Omega,\nonumber \\
\alpha & = & \frac{\Delta^{2}+\gamma^{2}}{(\Delta^{2}+\gamma^{2}-2\Gamma^{2})\gamma}\Omega,\label{eq:Steady-state solutions}\\
\beta_{2} & = & \frac{-\Gamma(i\Delta+\gamma)}{(\Delta^{2}+\gamma^{2}-2\Gamma^{2})\gamma}\Omega,\nonumber 
\end{eqnarray}
where the linear response will diverge under the condition $\Delta^{2}+\gamma^{2}-2\Gamma^{2}\rightarrow0$.
It is worth mentioning that the condition for linewidth suppression
(i.e. $E_{p}\rightarrow0$) is consistent with the divergent behavior
of the steady-state solutions in Eq. (\ref{eq:Steady-state solutions}).
This implies that linewidth suppression may lead to a considerable
sensing potential for internal or external perturbations \citep{PhysRevLett.126.180401}.
In the following, we illustrate that the scheme can be used to sense
optical Kerr-type nonlinearities, which arise from the nonlinear response
of the electric polarization in optical resonators, as well as to
detect intrinsic magnonic nonlinearities originating from nonlinear
magnetization dynamics. Such nonlinearities are ubiquitous in dispersive
atom--resonator \citep{Haroche2006,PhysRevA.76.042319}, magnon--photon
\citep{ZARERAMESHTI20221,PhysRevLett.121.137203}, and optomechanical
interactions \citep{RevModPhys.86.1391}. Remarkably, recent experiments
have demonstrated substantial Kerr coefficients in optical cavities
($U_{a}/2\pi=-12.2\pm0.1$ kHz/photon \citep{PhysRevLett.130.033601}).
In YIG spheres, the Kerr coefficient can be tuned from 0.05 to 100
nHz as the sphere diameter is varied from 1 mm to 100 $\mu$m \citep{PhysRevB.94.224410,PhysRevLett.120.057202}.

\subsection{\label{sectionIII A}Nonlinear cavity}

To explore nonlinearity sensing, we begin by introducing the nonlinear
term associated with the cavity mode. The modified steady-state equations
for the magnon and cavity modes are 
\begin{eqnarray}
0 & = & -i(\Delta-i\gamma)\beta_{1}-\Gamma\alpha,\nonumber \\
0 & = & -\gamma\alpha-2iU_{a}|\alpha|^{2}\alpha-\Gamma\beta_{1}-\Gamma\beta_{2}+\Omega,\label{eq:Steady-state relations-1}\\
0 & = & -i(-\Delta-i\gamma)\beta_{2}-\Gamma\alpha.\nonumber 
\end{eqnarray}
By eliminating $\beta_{1}$ and $\beta_{2}$, the cavity response
$x=|\alpha|^{2}$ is found to satisfy a cubic relation

\begin{eqnarray}
I & = & 4U_{a}^{2}x^{3}+\frac{\gamma^{2}E_{p}^{2}}{(\Delta^{2}+\gamma^{2})^{2}}x,\label{eq:CubicRelation}
\end{eqnarray}
where $I=\Omega^{2}$. The optical bistability can not occur since
the derivative $dI/dx=0$ has no solution for any drive power, see
Appendix B. This represents another advantage over the two-mode anti-PT
symmetric system, in which bistability would be observed when the
laser power $I$ is greater than a threshold value \citep{PhysRevB.103.224401}.
In the vicinity of linewidth suppression points, i.e. $E_{p}\rightarrow0$,
the linear term in Eq. (\ref{eq:CubicRelation}) becomes negligible,
and the cavity response approximates to $x\approx(I/4U_{a}^{2})^{\frac{1}{3}}$
. Note that $x$ is highly sensitive to variations in $U$, and the
sensitivity to $U_{a}$ in the response goes as $|\partial x/\partial U_{a}|\propto|U_{a}|^{-\frac{5}{3}}$.
It is worthwhile to emphasize that the inherent damping for the magnon
modes, which is inevitable due to mode defects, would reduce the sensitivity
to nonlinearities as the linewidth suppression condition can not be
met with the conventional two-lossy-mode setup \citep{PhysRevLett.126.180401}.
In contrast, the three-mode setup with an active cavity helps sustain
the high sensitivity.

In Fig. \ref{Response1}(a), we plot the response $x$ as a function
of $\Delta$ and $\gamma$ with the nonlinearity strength $U_{a}/2\pi=1\textrm{ nHz}$
and a weak drive power $P_{1,in}=8\textrm{ \ensuremath{\mu}W}$. As
illustrated by the dashed line in Fig. \ref{Response1}(a), for a
fixed value of $\gamma$, there may exist two detunings $\Delta$
corresponding to linewidth suppression points, where the cavity displays
a strong and sharply varying response near the suppression points,
enabling effective sensing while avoiding any instability. Moreover,
in Fig. \ref{Response1}(b), we show the line cuts along $\gamma=\sqrt{2}\Gamma$
for two different nonlinearities $U_{a1}/2\pi=0.1\textrm{ nHz}$ and
$U_{a2}/2\pi=1\textrm{ nHz}$, respectively. The peak response is
found at resonance $\Delta=0$ as a result of linewidth suppression.
In analogy to the lossless two-mode system \citep{PhysRevLett.126.180401},
a tenfold decrease in nonlinearity strength gives rise to a 4.64-fold
enhancement of the cavity response, hence leading to a remarkably
enhanced sensitivity to nonlinearities. However, in the above case
with $\gamma=\sqrt{2}\Gamma$, the decay rates of the two magnon modes
are $\gamma_{0}=(\sqrt{2}-1)\Gamma$, implying that the sensing of
weak optical nonlinearities on the order of $\textrm{nHz}$ can be
achieved even in the presence of two highly lossy magnon modes. In
addition, we increase the input power to $P_{2,in}=8\textrm{ mW}$,
which broadens the lineshape of the cavity response without changing
the ratio $x(U_{a1})/x(U_{a2})$ at $\Delta=0$. Importantly, under
this condition the system is stable for all detunings $\Delta$, and
Fig. \ref{Response1}(b) shows that the response remains robust against
small detuning fluctuations. This ensures that the operating point
is experimentally accessible and insensitive to small noise in $\Delta$.

So far, we consider that the laser driving is in resonance with the
cavity mode $a$. However, we find that an appropriate detuning $\Delta_{a}$
for the cavity mode $a$ can further enhance the sensitivity outside
the optical bistability regime. For this purpose, we quantify the
sensitivity with the measure defined by the ratio $\eta=x(U_{a1})/x(U_{a2})$,
where $x(U_{a1})$ and $x(U_{a2})$ denote the responses corresponding
to cavity nonlinearity strengths $U_{a1}=0.1\textrm{ \ensuremath{\mu}Hz}$
and $U_{a2}=1\textrm{\ensuremath{} \ensuremath{\mu}Hz}$, respectively.
As shown in Fig. \ref{Response1}(c), we plot $\eta$ as a function
of the detuning $\Delta_{a}$, where the solid and the dotted lines
denote the monostable and bistable regimes, respectively, see Appendix
B. Notably, when the detuning satisfies $0.08\Gamma>\Delta_{a}>0$,
the ratio $\eta\sim7$ can be achieved in the monostable regime. This
ratio is much greater than the typical 4.64-fold enhancement at the
linewidth suppression point.

\begin{figure}
\includegraphics[width=1\columnwidth]{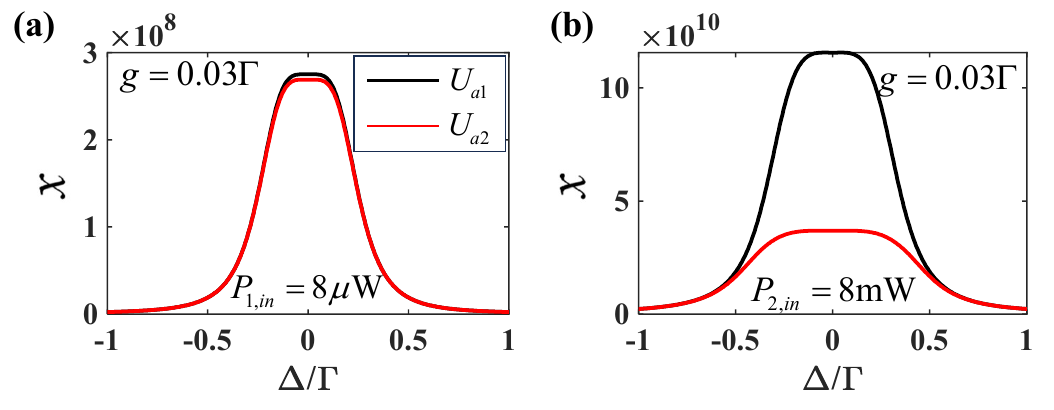}

\caption{\label{Resistance} The cavity responses plotted against $\Delta$
in the presence of a coherent coupling $g=0.03\Gamma$ with drive
powers (a) $8\textrm{ \ensuremath{\mu}W}$ and (b) $8\textrm{ mW}$.
Other parameters are the same as in Fig. \ref{Response1}(c).}
\end{figure}

In addition, we consider the effect of a finite coherent coupling
$g$ between the modes with $\Delta_{a}=0$. For $g\neq0$, the real
singularity is now replaced by a complex one, i.e., the eigenvalue
$\lambda_{1}=-i\gamma+i\sqrt{2\Gamma^{2}+4ig\Gamma-2g^{2}-\Delta^{2}}$
has a nonvanishing imaginary part with $E_{p}\rightarrow0$. In this
case, a tenfold variation in $U_{a}$ can only lead to a small change
in cavity response as shown in Fig. \ref{Resistance}(a). Thus, the
sensitivity becomes weak as $\eta\sim1$. Nevertheless, by increasing
the strength of input laser power from $8\textrm{ \ensuremath{\mu}W}$
to $8\textrm{ mW}$, one can again recover high sensitivity, see Fig.
\ref{Resistance}(b).

\subsection{\label{sectionIII B}Nonlinear magnonic system}

We now turn to sensing the nonlinearity in one of the magnon modes
and suppose a nonlinear $b_{1}$ without loss of generality. The steady-state
relations are now given by 
\begin{eqnarray}
0 & = & -i(\Delta-i\gamma)\beta_{1}-2iU_{b_{1}}|\beta_{1}|^{2}\beta_{1}-\Gamma\alpha,\nonumber \\
0 & = & -\gamma\alpha-\Gamma\beta_{1}-\Gamma\beta_{2}+\Omega,\label{eq:Steady-state relations2-1}\\
0 & = & -i(-\Delta-i\gamma)\beta_{2}-\Gamma\alpha.\nonumber 
\end{eqnarray}
From these, we obtain the spin-current response of the YIG sphere, defined as $y=|\beta_{1}|^{2}$, which satisfies the cubic relation:
\begin{eqnarray}
I & = & \frac{4U_{b_{1}}^{2}[(\gamma^{2}-\Gamma^{2})^{2}+\gamma^{2}\Delta^{2}]y^{3}-4U_{b_{1}}\Delta\gamma^{2}E_{p}y^{2}+\gamma^{2}E_{p}^{2}y}{\Gamma^{2}(\Delta^{2}+\gamma^{2})}.\nonumber \\
\end{eqnarray}
In the vicinity of the linewidth suppression points where $E_{p}\rightarrow0$,
the coefficients of the quadratic term $y^{2}$ and the linear term
$y$ become vanishingly small. As a result, the spin-current response
becomes highly sensitive to variations in $U_{b_{1}}$ and approximately
follows the functional dependence 
\begin{eqnarray}
y & \approx & \left\{ \frac{I\Gamma^{2}(\Delta^{2}+\gamma^{2})}{4U_{b_{1}}^{2}[(\gamma^{2}-\Gamma^{2})^{2}+\gamma^{2}\Delta^{2}]}\right\} {}^{\frac{1}{3}}\approx\left(\frac{I}{2U_{b_{1}}^{2}}\right){}^{\frac{1}{3}}.\nonumber \\
\end{eqnarray}
Hence, the sensitivity to $U_{b_{1}}$ in the response is again encoded
as $\left|\partial y/\partial U_{b_{1}}\right|\propto|U_{b_{1}}|^{-\frac{5}{3}},$
which is the same as the case of a nonlinear cavity. Alternatively,
based on Eq. (\ref{eq:Steady-state relations2-1}), the cavity response
$x=|\alpha|^{2}$ can also be written by 
\begin{equation}
I=\left|\frac{\Gamma^{2}}{-i\left(\Delta+2U_{b_{1}}y\right)-\gamma}+\gamma+\frac{\Gamma^{2}}{i\Delta-\gamma}\right|^{2}x.
\end{equation}
For $E_{p}\rightarrow0$ and considering the EP singularity with respect
to $\Delta=0$, one can again find that $\left|\partial x/\partial U_{b_{1}}\right|\sim\frac{\partial}{\partial U}\left(I/U_{b_{1}}^{2}\right)^{1/3}\propto|U_{b_{1}}|^{-\frac{5}{3}}$.
Without loss of generality, we show the spin current response in what
follows.

\begin{figure}
\includegraphics[width=1\columnwidth]{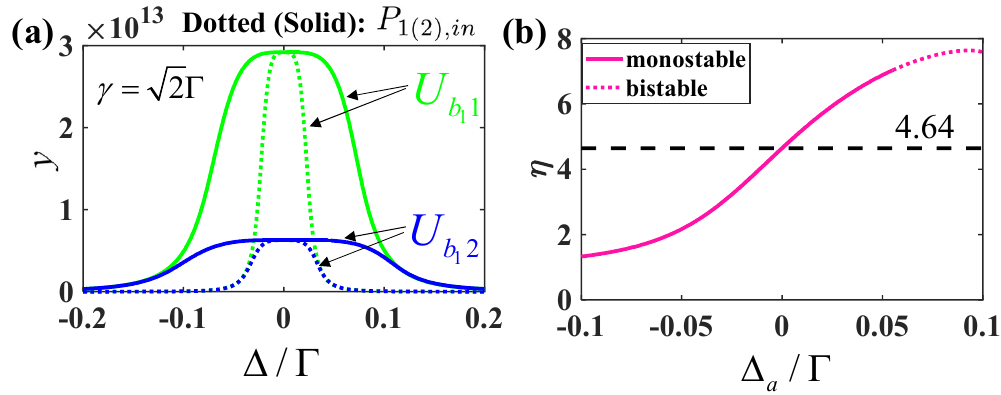}

\caption{\label{Response2} (a) The magnon spin-current responses plotted against
$\Delta$ for $\gamma=\sqrt{2}\Gamma$ with weak Kerr nonlinearity
strengths $U_{b_{1}1}/2\pi=0.1\textrm{ nHz}$ and $U_{b_{1}2}/2\pi=1\textrm{ nHz}$.
The green- and blue-dotted curves have been scaled up by 10. The spin-current
responses show similar behavior to those in sensing weak cavity nonlinearity.
(b) $\eta$ as a function of $\Delta_{a}$, at the drive power of
$8\textrm{ mW}$ for two different nonlinearity strengths $U_{b_{1}1}=0.1\textrm{ \ensuremath{\mu}Hz}$
and $U_{b_{1}2}=1\textrm{ \ensuremath{\mu}Hz}$. The monostable regime
and bistable regime are denoted by the pink solid line and dotted
line, respectively. The horizontal dashed line represents the response
enhancement factor of 4.64, corresponding to the sensitivity at the
linewidth suppression condition. Other parameters are the same as
Fig. \ref{Response1}(c).}
\end{figure}

To demonstrate the sensing of Kerr nonlinearities intrinsic to magnon
mode, we examine the response for two weak nonlinearity strengths
$U_{b_{1}}=0.1\textrm{ nHz}$ and $U_{b_{1}}=1\textrm{ nHz}$ under
the weak drive power $P_{in}=8\textrm{ \ensuremath{\mu}W}$. As shown
in Fig. \ref{Response2}(a), we plot the response $y$ against $\Delta$
for the intrinsic decay rates of the magnonic modes being $\gamma_{b_{j}}=\left(\sqrt{2}-1\right)\Gamma$
and the optical gain $-\gamma_{a}=(2-\sqrt{2})\Gamma$. For $\Delta=0$,
a tenfold decrease in nonlinearity strength again gives rise to a
4.64-fold enhancement of the spin current response as a result of
the linewidth suppression. Furthermore, in Fig. \ref{Response2}(b),
we show the ratio of the spin current response $\eta=y(U_{b_{1}1})/y(U_{b_{1}2})$
by introducing an appropriate detuning $\Delta_{a}$ from cavity resonance,
with $U_{b_{1}1}=0.1\textrm{ \ensuremath{\mu}Hz}$ and $U_{b_{1}2}=1\textrm{ \ensuremath{\mu}Hz}$.
As the system keeps away from the bistable regime (the dotted part),
a higher sensitivity of $\eta\sim7$ can be reached for $\Delta_{a}\sim0.05\Gamma$,
analogous to what is observed in a nonlinear cavity.

\section{\label{section IV}Experimental realization and discussion}

To construct a three-mode passive-active anti-PT symmetric system,
the cavity mode must be active, which can be achieved by adding an
external driving field on auxiliary qubits \citep{Quijandria2018}
or optical gain medium \citep{Zhang2025}. The damping rates and resonance
frequencies of the magnon modes, which arise from the surface roughness,
the impurities, and defects in the YIG sphere, can be alternatively
controlled by a grounded loop antenna getting close to the YIG sphere
and an external magnetic field \citep{PhysRevLett.125.147202}. With
that in hand, we consider an integrated apparatus comprising two YIG
spheres and a microwave cavity, where the cavity is dissipatively
coupled to the two YIG spheres through two different waveguides. Owing
to the nonexistent spatial overlap between the cavity and magnon modes,
the direct couplings between them are dropped. Then the full Hamiltonian
in the presence of the external drive can be cast exactly in the form
of Eq. (\ref{eq:Hamiltonian}). By tuning the damping and gain of
the modes, the passive-active anti-PT symmetric system can be realized
\citep{PhysRevLett.121.137203,PhysRevB.99.134426}.

The three-mode anti-PT-symmetric configuration is well suited for
detecting weak anharmonicities associated with passive magnon modes,
the detectable range of the anharmonicity ($U_{a}$ or $U_{b_{j}}$)
is determined by the regime in which the nonlinear correction remains
weak (i.e., $\{U_{a}|\alpha|^{2},U_{b_{j}}|\beta|^{2}\}/\Gamma<1$),
so that the linear coupling between the cavity and magnonic modes
continues to dominate the response. For anharmonicities $\{U_a,U_{b_{j}}\}$ in the nHz ($\mu$Hz) range, the corresponding ratios are on the order of $10^{-3}$ ($10^{-1}$).
In these weakly nonlinear regimes,
the steady-state solution reduces to $x\simeq(I/4U_{a}^{2})^{1/3}$
{[}or $y\simeq(I/4U_{b_{j}}^{2})^{1/3}${]}, implying a sensitivity
$\left|\partial x/\partial U_{a}\right|\propto U_{a}^{-5/3}$ {[}or
$\left|\partial y/\partial U_{b_{j}}\right|\propto U_{b_{j}}^{-5/3}${]},
which diverges as $U_{a(b_{j})}\rightarrow0$. Thus, an extremely
small nonlinearity, down to the nHz scale in the cavity or YIG implementation,
remain detectable. The practical upper bound on $U_{a(b_{j})}$ is
set by the onset of strong-drive nonlinear effects when $\{U_{a}x,U_{b_{j}}y\}\sim\Gamma$,
where higher-order terms modify the energy spectrum of the anti-PT
symmetric system and degrade the linewidth suppression induced sensitivity
enhancement. At low powers the response obeys $\{x,y\}\sim I^{1/3}$,
so the signal increases with drive but without entering instability.
When the drive is too strong, additional nonlinear resonances appear,
reducing the monotonic sensitivity to $U_{a(b_{j})}$. Similarly,
the sensitivity increases when $\Delta_{a}$ is tuned towards the
onset of the bistability. Beyond this point, $\{U_{a}x,U_{b_{j}}y\}$
may undergo a transition and exhibit nonlinear behavior - a regime
in which the sensing scheme becomes ineffective.

In summary, we have investigated the enhanced sensing of weak nonlinearities
within a passive-active anti-PT symmetric system where the active
cavity mode dissipatively couples to two magnon modes with intrinsic
losses. The system Hamiltonian maintains the anti-PT symmetry characteristic,
and its eigenvalue may display linewidth suppression points, which
can be actively controlled by the optical gain even when the magnon
modes experience independent losses. The EP-like singularity associated
with linewidth suppression has been utilized to detect weak nonlinearities
in both the cavity mode and the magnonic modes \citep{PhysRevLett.126.180401}.
For the scenario where only the cavity mode exhibits nonlinearity,
a tenfold decrease in nonlinearity strength gives rise to a 4.64-fold
enhancement of the cavity response with a resonant cavity driving,
where optical bistability can not happen, regardless of the driving
strength and the nonlinearities. An appropriate drive detuning to
cavity mode helps to increase and sensitivity by about fifty percent.
For the case where one of the magnon modes contains Kerr nonlinearity,
we obtain the same enhanced sensitivity manifested in both the cavity
and spin current responses. Finally, we emphasize that the scheme
does not necessitate the intrinsic losses of the modes to be zero
and can be extended to a wide variety of systems, such as laser-cooled
atomic ensembles \citep{PhysRevLett.117.173602}, superconducting
transmon qubits \citep{doi:10.1126/science.aaa3693,https://doi.org/10.1002/qute.202300350,PhysRevLett.125.117701},
and optomechanical systems \citep{fang2016optical,Harris2016Topological,PhysRevLett.108.120801,Xiong2021}.
\begin{acknowledgments}
We thank Yong Li for helpful discussions. H.W. acknowledges support
from the National Natural Science Foundation of China under Grant
No. 12174058.
\end{acknowledgments}

\appendix

\section{\label{Appendix A}Stability of the system}

A coupled three-mode system with nonlinearities may exhibit instability
under a specific drive power, and such instability is generally deemed
undesirable. To ensure the system operates within a stable regime,
we turn to the linearized dynamics. As presented in the main text,
the system dynamics, expressed in terms of the annihilation (creation)
operators, can be expanded as follows: 
\begin{eqnarray}
\dot{b_{1}} & = & -i\Delta_{1}b_{1}-\gamma_{b_{1}}b_{1}-\Gamma b_{1}-\Gamma a+b_{1,in},\nonumber \\
\dot{b_{1}^{\dagger}} & = & i\Delta_{1}b_{1}^{\dagger}-\gamma_{b_{1}}b_{1}^{\dagger}-\Gamma b_{1}^{\dagger}-\Gamma a^{\dagger}+b_{1,in}^{\dagger},\nonumber \\
\dot{a} & = & -i\Delta_{a}a-\gamma_{a}b-2iU_{a}a^{\dagger}aa-2\Gamma a\nonumber \\
 &  & -\Gamma b_{1}-\Gamma b_{2}+\Omega+a_{in},\nonumber \\
\dot{a^{\dagger}} & = & i\Delta_{a}a^{\dagger}+2iU_{a}a^{\dagger}a^{\dagger}a-2\Gamma a^{\dagger}\nonumber \\
 &  & -\Gamma b_{1}^{\dagger}-\Gamma b_{2}^{\dagger}+\Omega+a_{in}^{\dagger},\nonumber \\
\dot{b_{2}} & = & -i\Delta_{2}b_{2}-\gamma_{b_{2}}b_{2}-\Gamma b_{2}-\Gamma a+b_{2,in},\nonumber \\
\dot{b_{2}^{\dagger}} & = & i\Delta_{2}b_{2}^{\dagger}-\gamma_{b_{2}}b_{2}^{\dagger}-\Gamma b_{2}^{\dagger}-\Gamma a^{\dagger}+b_{2,in}^{\dagger}.
\end{eqnarray}
where $a_{in}$, $b_{1,in}$, $b_{1,in}$ are white Gaussian noise
with zero mean. Considering the anti-PT symmetric configuration, we
impose the following conditions: $\Delta=\Delta_{1}=-\Delta_{2}$,
$\Delta_{a}=0$, $\gamma_{b_{1}}=\gamma_{b_{2}}=\gamma_{0}$ and $\gamma=\gamma_{0}+\Gamma=\gamma_{a}+2\Gamma$.
To incorporate fluctuations, we linearize the operators around the
classical mean values as $b_{1}=\beta_{1}+\delta b_{1}$, $b_{2}=\beta_{2}+\delta b_{2}$
and $a=\alpha+\delta a$. The dynamics of the fluctuations can be
further formulated as (by omitting quantum noise terms): 
\begin{eqnarray}
\dot{\delta b_{1}} & = & (-i\Delta-\gamma)\delta b_{1}-\Gamma\delta a,\nonumber \\
\dot{\delta b_{1}^{\dagger}} & = & (i\Delta-\gamma)\delta b_{1}^{\dagger}-\Gamma\delta a^{\dagger},\nonumber \\
\dot{\delta a} & = & -\gamma\delta a-2iU_{a}(2|\alpha|^{2}\delta a+\alpha^{2}\delta a^{\dagger})-\Gamma\delta b_{1}-\Gamma\delta b_{2},\nonumber \\
\dot{\delta a^{\dagger}} & = & -\gamma\delta a^{\dagger}+2iU_{a}(2|\alpha|^{2}\delta a^{\dagger}+\alpha^{*2}\delta a)-\Gamma\delta b_{1}^{\dagger}-\Gamma\delta b_{2}^{\dagger},\nonumber \\
\dot{\delta b_{2}} & = & (i\Delta-\gamma)\delta b_{2}-\Gamma\delta a,\nonumber \\
\dot{\delta b_{2}^{\dagger}} & = & (-i\Delta-\gamma)\delta b_{2}^{\dagger}-\Gamma\delta a^{\dagger}.
\end{eqnarray}
At a low drive power, the dynamics of the system can be characterized
by a 3 by 3 anti-PT symmetric matrix, as the product $U|\alpha|^{2}$
is significantly smaller than $\Gamma$. The property of the system
is predominantly determined by the eigenvalues of the Hamiltonian
$H$ described in the main text. This characteristic is attributed
to the modified spectroscopic characteristics of the system under
a weak probe. However, as the drive power increases, the nonlinear
optical shift $U|\alpha|^{2}$ becomes comparable to $\Gamma$ or
even much larger than $\Gamma$. Thus, a more accurate description
necessitates employing the higher-order matrix.

To inspect the system stability, the quadratures are utilized instead
to describe the dynamics of the system: $\delta q_{b_{1}}=(\delta b_{1}+\delta b_{1}^{\dagger})/\sqrt{2}$,
$\delta p_{b_{1}}=(\delta b_{1}-\delta b_{1}^{\dagger})/i\sqrt{2}$,
$\delta q_{a}=(\delta a+\delta a^{\dagger})/\sqrt{2}$, $\delta p_{a}=(\delta a-\delta a^{\dagger})/i\sqrt{2}$,
$\delta q_{b_{2}}=(\delta b_{2}+\delta b_{2}^{\dagger})/\sqrt{2}$,
$\delta p_{b_{2}}=(\delta b_{2}-\delta b_{2}^{\dagger})/i\sqrt{2}$.
For the new basis of the quadratures $\delta\varsigma=[\delta q_{b_{1}},\delta p_{b_{1}},\delta q_{a},\delta p_{a},\delta q_{b_{2}},\delta p_{b_{2}}]$,
the dynamics is dominated by $\text{d\ensuremath{\delta\varsigma}/d}t=\mathcal{M}\delta\varsigma$
with
\begin{widetext}
\begin{eqnarray}
\mathcal{M} & = & \left(\begin{array}{cccccc}
-\gamma & \Delta & -\Gamma & 0 & 0 & 0\\
-\Delta & -\gamma & 0 & -\Gamma & 0 & 0\\
-\Gamma & 0 & -\gamma-i\varLambda & \Sigma-\Xi & -\Gamma & 0\\
0 & -\Gamma & -\Sigma-\Xi & -\gamma+i\varLambda & 0 & -\Gamma\\
0 & 0 & -\Gamma & 0 & -\gamma & -\Delta\\
0 & 0 & 0 & -\Gamma & \Delta & -\gamma
\end{array}\right),
\end{eqnarray}
where $\varLambda=U_{a}\left(\alpha^{2}-\alpha^{*2}\right)$, $\Xi=U_{a}\left(\alpha^{2}+\alpha^{*2}\right)$,
and $\Sigma=4U_{a}|\alpha|^{2}$. Following a consideration of the
Routh-Hurwitz criterion \citep{PhysRevA.35.5288}, the real parts
of the eigenvalues of the matrix $\mathcal{M}$ must be strictly negative
to ensure that the system is stable.
\end{widetext}

\section{\label{Appendix B} optical bistability for $\Delta_{a}\protect\neq0$}

For $\Delta_{a}\neq0$ and $g=0$, the steady-state relations of the
three-mode system are 
\begin{eqnarray}
0 & = & -i(\Delta-i\gamma)\beta_{1}-\Gamma\alpha,\nonumber \\
0 & = & -i(\Delta_{a}-i\gamma)\alpha-\Gamma\beta_{1}-\Gamma\beta_{2}\nonumber \\
 &  & +2iU_{a}|\alpha|^{2}\alpha+\Omega,\\
0 & = & -i(-\Delta-i\gamma)\beta_{2}-\Gamma\alpha.\nonumber 
\end{eqnarray}
Eliminating $\beta_{1}$ and $\beta_{2}$, the cubic relation for
the cavity response takes the form as 
\begin{eqnarray}
I & = & 4U_{a}^{2}x^{3}-4U_{a}\Delta_{a}x^{2}\nonumber \\
 &  & +[\frac{\gamma^{2}(\Delta^{2}+\gamma^{2}-2\Gamma^{2})^{2}}{(\Delta^{2}+\gamma^{2})^{2}}+\Delta_{a}^{2}]x,
\end{eqnarray}
where the quadratic term $\sim x^{2}$ is nonvanishing, and may lead
to optical bistability.

\bibliographystyle{apsrev4-1}
\bibliography{Refs}
\end{document}